\documentclass[twocolumn,letterpaper,aps,prl,showpacs,floatfix,10pt]{revtex4-1}
\usepackage{graphicx}    
\usepackage{hyperref}    
\usepackage{bm}          
\usepackage[sumlimits,intlimits]{amsmath}
\usepackage{amsfonts,amssymb}

\pdfoutput=1

\newcommand{\be}{\begin{equation}}
\newcommand{\ee}{\end{equation}}

\begin{document}

\title{A simple variational method for calculating energy and quantum capacitance of an electron gas with screened interactions}

\date{\today}

\author{Brian Skinner}
\affiliation{Fine Theoretical Physics Institute, University of Minnesota, Minneapolis, Minnesota 55455}

\author{Michael M. Fogler}
\affiliation{Department of Physics, University of California--San Diego, 9500 Gilman Drive, La Jolla, California 92093}

\begin{abstract}

We describe a variational procedure for calculating the energy of an electron gas in which the long-range Coulomb interaction is truncated by the screening effect of a nearby metallic gate. We use this procedure to compute the quantum capacitance of the system as a function of electron density and spin polarization. The accuracy of the method is verified against published Monte-Carlo data. The results compare favorably with a recent experiment.

\end{abstract}
\maketitle

The electron gas with $1/r$ Coulomb interactions is a fundamental reference system. It provides both a testing ground for many-body theory and a microscopic input for the density functional theory (DFT), which is the main tool of electron structure calculations~\cite{Giuliani2005qto}. However, there are compelling reasons to study electrons with interactions other than Coulomb. One is a recent interest in mixed DFT schemes, where the Coulomb potential is split into a long-range part, to be handled by other techniques, and a short-range part, to be treated within the usual DFT. This allows one to capture long-range van~der~Waals interaction effects but necessitates recomputing the exchange-correlation energies for the truncated Coulomb potential~\cite{Zecca2004ldf}. Another and more direct motivation comes from the physics of low-dimensional systems, in which the bare interaction is often modified by the environment.  One example is a two-dimensional (2D) electron gas positioned a small distance $d$ away from a metallic gate. The gate creates an image charge for each electron, leading to the interaction law
\begin{equation} 
V(r) = \frac{e^2}{\kappa r} - \frac{e^2}{\kappa \sqrt{r^2 + 4 d^2}}\,,
\label{eqn:Vr}
\end{equation}
where $\kappa$ is the dielectric constant of the medium. At $r \gg d$, this potential rapidly decays:
$V(r) \simeq 2 e^2 d^2/\kappa r^3$. Therefore, electron correlations at low density $n \ll 1/d^2$ are very different from those in the absence of the gate. Previously, these correlations have been treated using a semi-classical theory in which electrons are assumed to form a crystal~\cite{Widom1988cea, Skinner2010alc}. Such an approximation underestimates the energy but it is legitimate in the range of densities $a^2 / d^4 \ll n \ll 1 / a^2$, where $a = \hbar^2 \kappa / m e^2$ is the effective Bohr radius. Outside of this range, a different approach is needed.

Generally speaking, the most reliable results for the electron gas have been obtained by quantum Monte Carlo (MC) simulations. While some non-Coulomb potentials have been examined~\cite{Ceperley1977mcs} in applications to $^3\text{He}$ and nuclear matter in three dimensions (3D), no simulations for the interaction law $V(r)$ in 2D have been reported. Below we demonstrate a variational method for calculating the total energy $E$ per electron, which is accurate yet simple to implement for any truncated Coulomb potential. Our approach is to treat the electron charge $e$ in $e^2 / r$ as an adjustable parameter and use the corresponding ground states as trial states for the system with the desired interaction law $V(r)$. Since the $1/r$ interaction does not have an intrinsic lengthscale, the variational energies can be obtained from the already existing Monte Carlo data by a suitable rescaling.

%
%
\begin{figure}[b]
\centering
\includegraphics[width=0.45 \textwidth]{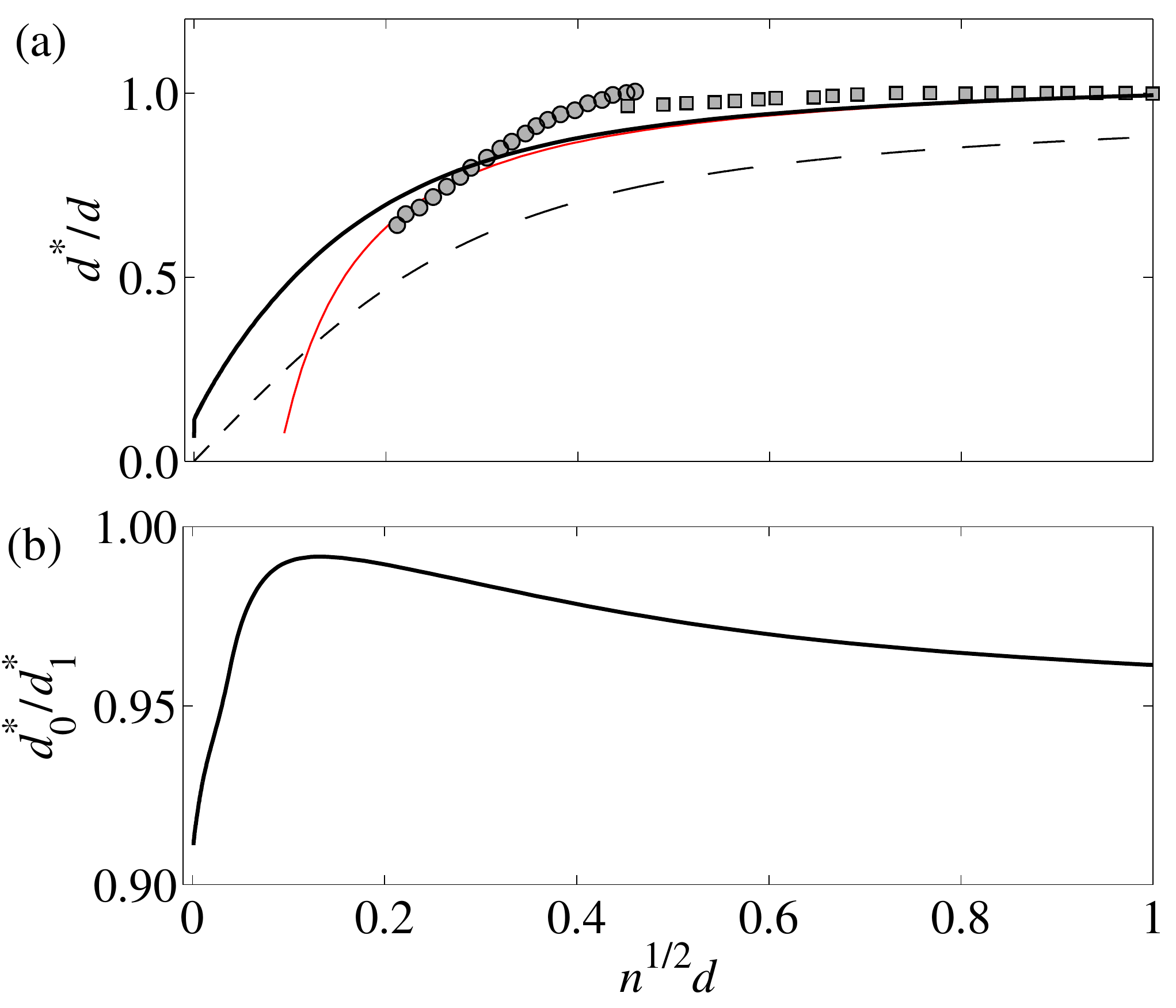}
\caption{(Color online) (a) The effective capacitor thickness $d^*$ in units of $d$ as a function of $n^{1/2}d$ for the spin unpolarized case, with and without the gate screening effect (thick black and thin red lines, respectively). The dashed line is the previous classical theory~\cite{Skinner2010alc}. Symbols are the experimental data~\cite{Li2010lce}. (b) The ratio of $d^*$ for the spin unpolarized ($\zeta = 0$) and polarized ($\zeta = 1$) cases.} \label{fig:dstar}
\end{figure}

Experimentally, the regime $n d^2 < 1$ has been probed in several recent studies~\cite{Li2010lce, Xia2009mot, Young2010eco, Ponomarenko2010dos}. Our theory can be compared with the one~\cite{Li2010lce} done on a $\text{LaAlO}_3/\text{SrTiO}_3$ heterostructure with $d = 4\,\text{nm}$. In Fig.~\ref{fig:dstar}(a) we present the results for the effective capacitor thickness $d^*$, which is related to $E(n)$ by
\begin{equation} 
d^* = \frac{\kappa}{4\pi e^2}\, \frac{d\mu}{d n} = \frac{\kappa}{4 \pi e^2}\, \frac{d^2}{d n^2} \bigl(E(n) n \bigr)\, ,
\label{eqn:dstar}
\end{equation} 
where $\mu(n)$ is the electrochemical potential.  This quantity $d^*$ can be extracted from the measured differential capacitance of the electron gas. Our results are in a good agreement with the experiment~\cite{Li2010lce}; however, the gate screening effect becomes important only at the lowest measured densities [see Fig.~\ref{fig:dstar}(a)]. Below we first describe how these results have been obtained and then show that our method captures the essential physics of the problem at both low and high densities.  Next, we check that our solution satisfies the virial theorem and verify the accuracy of our method on a 3D example. Finally, we discuss experimental implications.

\textit{Variational procedure.} \;
The trial states for our variational calculation are the ground-states of the 2D electron gas with the $e^2/r$ interaction on a neutralizing background. The total energy per electron $E$, the interaction energy $E^{\text{(int)}}$, and the pair distribution function (PDF) $g(r)$ of this system are known to scale with $n$ and with the dimensionless parameter $r_s = 1 / \sqrt{\pi n a^2}$. For example,
\begin{equation}
E = E_0 f(r_s)\,,
\quad
E^{\text{(int)}} = E_0 f^{\text{(int)}}(r_s)
\quad (\text{no gate})\,.
\label{eqn:f}
\end{equation}
Here $E_0 = \pi \hbar^2 n / m$ is the Fermi energy of a noninteracting 2D Fermi gas and  $f(r_s)$ is a dimensionless function~\cite{Giuliani2005qto}. By the virial theorem, $f^{\text{(int)}}(r_s) = r_s f^\prime(r_s)$. Similarly, we can write $g(r) = G(k_F r, r_s)$, where $G$ is another dimensionless function and $k_F = \sqrt{2 \pi n}$ is the Fermi momentum. We rely on the fact that $f$ and $G$ have been computed by MC techniques and fitted to analytical expressions~\cite{Tanatar1989gso, Attaccalite2002cea, Gori-Giorgi2004pdf} over the broad range $1 \leq r_s \leq 40$. This enables us to use $e^2$, or equivalently, $r_s$, as a variational parameter that labels the trial wavefunction. We denote this parameter $\rho_s$ to distinguish it from the physical $r_s$. Large $\rho_s$ describes a strongly-correlated, crystal-like arrangement of electrons~\cite{Widom1988cea, Skinner2010alc}, while $\rho_s = 0$ corresponds to the free Fermi gas, as in the Hartree-Fock (HF) approximation. The optimal $\rho_s$ is to be found by minimizing the variational energy $E_{\text{var}}(n, \rho_s) = E_{\text{var}}^{(\text{kin})}(n, \rho_s) + E_{\text{var}}^{\text{(int)}}(n, \rho_s)$. We expect $\rho_s < r_s$ because screening by the gate reduces electron repulsion and weakens electrostatic correlations.

%
%
\begin{figure}[t]
\centering
\includegraphics[width=0.45 \textwidth]{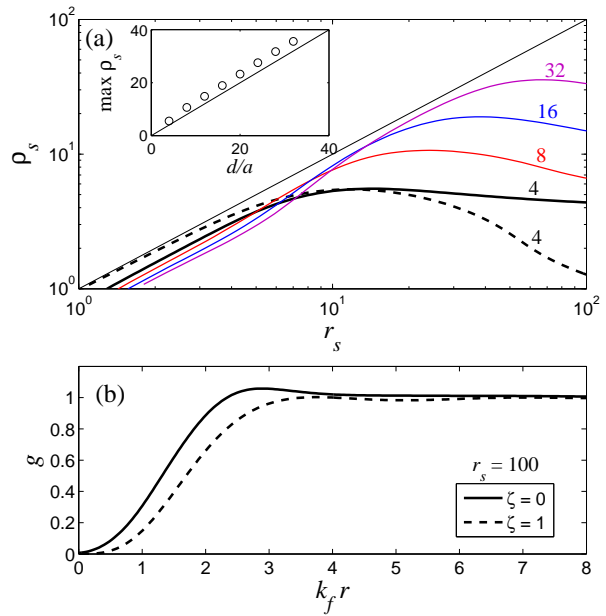}
\caption{(Color online) (a) Optimal $\rho_s$ \textit{vs.\/} $r_s$. Solid (dashed) lines correspond to $\zeta = 0$ ($\zeta = 1$). The labels indicate the values of $d/a$. The inset depicts the maximum value of $\rho_s$ for $\zeta = 0$ (circles) as a function of $d/a$ (line).
(b) Pair distribution functions $g(r)$ for the optimal trial states at $r_s = 100$.} \label{fig:rscompare}
\end{figure}

For a given trial $\rho_s$, the kinetic energy $E_{\text{var}}^{(\text{kin})}(n, \rho_s)$ is obtained by subtracting the interaction energy from the total energy for the $1/r$ interaction:
\begin{equation} 
E_{\text{var}}^{(\text{kin})}(n, \rho_s) = E_0\, \left[f(\rho_s) - \rho_s f^\prime(\rho_s)\right].
\label{eqn:ekin1}
\end{equation} 
By making use of the definition $E = E_0(1 + \zeta^2) / 2 + E_x + E_c$, where $E_x$ is the exchange energy, $E_c = E_0 f_c(r_s)$ is the correlation energy, and $0 \leq \zeta \leq 1$ is the spin polarization, we can rewrite Eq.~\eqref{eqn:ekin1} as
\begin{equation} 
E_{\text{var}}^{(\text{kin})} = E_0 \left(\frac{1 + \zeta^2}{2}
 - \rho_s f_c^\prime(\rho_s)
 \right).
\label{eqn:ekin2}
\end{equation}
The function $f_c(\rho_s)$ is taken from Ref.~\onlinecite{Attaccalite2002cea}. The interaction energy $E^{\text{(int)}}_{\text{var}}$ is computed by numerical quadrature of the potential $V(r)$ weighted by the PDF,
\begin{equation}
E^{\text{(int)}}_{\text{var}}(n, \rho_s) =\frac12 \int\limits_0^\infty V\left(\frac{x}{k_F}\right)
 G(x, \rho_s) x d x - \frac{e^2}{4 \kappa d}\,.
\label{eqn:epot2}
\end{equation}
(The last term accounts for the interaction of each electron with its image.) For $G$, we use the parametrization of Ref.~\onlinecite{Gori-Giorgi2004pdf}. The minimization of $E_{\text{var}}$ over $\rho_s$ is done numerically. Finally, $d^*$ is computed according to Eq.~\eqref{eqn:dstar}.

We performed these calculations for $d / a$ ranging from $4$ to $32$ and for two values of the spin polarization, $\zeta = 0$ (unpolarized) and $1$ (fully polarized)~\footnote{Parameter $\zeta$ enters implicitly in functions $f_c$ and $G$.}. The optimal $\rho_{s}$ as a function of the physical $r_s$ is shown in Fig.~\ref{fig:rscompare}(a) for $\zeta = 0, 1$. At large electron density (small $r_s$), $\rho_{s}$ remains close to $r_s$ because the average spacing between electrons is much smaller than $d$, so that the gate plays a minor role. As $r_s$ increases, $\rho_{s}$ attains a maximum value $\rho_{s} \approx d / a$, and then drops off. This is because at $r_s > d / a$ the interaction between electrons becomes effectively short-ranged, so that electrons lose their electrostatic correlations as the system becomes more dilute. For $\zeta = 1$, the drop of $\rho_s$ at $r_s > d / a$ is fairly rapid. In contrast, in the unpolarized case, $\rho_{s}$ exhibits a broad plateau before it also collapses at very high $r_s$ [far beyond the range shown in Fig.~\ref{fig:rscompare}(a)].

In the remainder of this paper we show that our results withstand several tests: (i) they correctly reproduce the asymptotic behavior at large and small $n$, (ii) they obey the virial theorem, and (iii) when generalized to 3D, they yield a good agreement with MC simulations~\cite{Zecca2004ldf}.

\textit{Low and high density asymptotics. } \;  
Because of its fast $1/r^3$ decay, $V(r)$ belongs to the universality class of short-range potentials. At low $n$, the energy per particle coincides in the first approximation with that of the free Fermi gas, except that it is shifted by $-e^2 / 4 \kappa d$, cf.~Eq.~\eqref{eqn:epot2}. This is reproduced by our calculation since $\rho_s(n \rightarrow 0) = 0$. This fact further implies that $d^*_0 \equiv d^*(\zeta = 0)$ and $d^*_1 \equiv d^*(\zeta = 1)$ tend to $a / 4$ and $a / 2$, respectively, as $n \to 0$ [see Fig.~\ref{fig:dstar}(a)].

The leading interaction correction to the free gas limit comes from two-body collisions, which are parametrized by the effective hard-core radius~\cite{Meyertholen2008bit} $b = 2 e^{2\gamma} d^2 / a$, where $\gamma \approx 0.5772$ is the Euler constant. In the absence of spin polarization ($\zeta = 0$), the perturbative result for the electrochemical potential reads~\cite{Engelbrecht1992lff}
$\mu = E_0 [1 + 2 v + 4 v^2 (1 - \ln 2)]$, where~\cite{Randeria1990sia} $1/v = -2 \gamma - \ln(\pi n b^2)$. Accordingly,
\begin{equation} 
d^*_0 \simeq \frac{a}{4} \left[1 + \frac{4}{\ln (a^2\, /\, n d^4) - 6.30} \right]\,.
\label{eqn:dstar0}
\end{equation}
This expression is expected to apply at $v \lesssim 1/2$, which corresponds to extraordinary low densities, e.g., $n a^2 < 10^{-6}$ for $d = 4a$. Indeed, our variational method shows that the departure of $d^*_0$ from its zero-density value of $a / 4$ develops abruptly as a function of $n$ within a narrow interval $n^{1/2} d < 10^{-3}$ [see Fig.~\ref{fig:dstar}(a)]. Of course, at such $n$ the energy of real electron systems is dominated by disorder~\cite{Allison2006tdo}, which we do not consider here.

The abrupt growth of $d^*_0$ caused by the logarithmic correction in Eq.~\eqref{eqn:dstar0} reflects the low-energy behavior of the $s$-wave scattering phase-shift in 2D. It can be contrasted with the gradual increase of $d^*_1$ for the polarized gas, where $p$-wave scattering dominates. Here the Pauli exclusion between like-spin electrons ensures that electron pairs do not approach each other too closely and the short-range repulsion is more easily satisfied. As a result, in the polarized gas, we have $d^*_1 = ({a} / {2}) [1 + \mathcal{O}(n b^2)]$.

We can also consider the opposite limit, $n \gg 1/d^2$, where to the leading order $\rho_s = r_s$ and so the kinetic energy is unaffected by the gate. The correction $\Delta E(n, d)$ to the total energy with respect to the reference system without the gate is determined by the interaction energy. To compute the latter, we rewrite Eq.~\eqref{eqn:epot2} as
\begin{equation}
E^{\text{(int)}}_{\text{var}} = \frac{n}{2} \int V(r)
 [g(r) - 1] d^2 r + \frac{2\pi e^2 d}{\kappa}\, n - \frac{e^2}{4 \kappa d}\,.
\label{eqn:epot3}
\end{equation}
The difference $g(r) - 1$ is appreciable only at $|r| \lesssim n^{-1/2}$ and gives $-1 / n$ when integrated over all $r$. Therefore, to the order $\mathcal{O}(1/d)$ we can set $V(r) = e^2 / (\kappa r) - e^2 / (2 \kappa d)$ in the integral, leading to $\Delta E = ({2\pi e^2 d} / {\kappa}) n$.
The corresponding correction to the electrochemical potential is
$\Delta \mu = ({4 \pi e^2 d}/{\kappa}) n$, which is the classical relation between the voltage $\mu / e$ and the charge density $e n$ of a parallel-plate capacitor of thickness $d$.

The parameter $d^*$, as defined in Eq.\ (\ref{eqn:dstar}), represents the effective capacitor thickness, with $C = e^2 d n / d \mu \equiv \kappa / (4 \pi d^*)$ being the capacitance per unit area. In general, $d^*$ differs from the geometric thickness $d$, which is often called the quantum capacitance (QC) effect~\cite*{[{See }][{ and references therein}]Kopp2009cot}. Our preceding calculation shows that at high density the gate can modify the QC only to order $1 / nd^2$. A simple expression for this correction can be derived at $n \gg 1/a^2$, where the HF approximation applies. For $\zeta = 0$, we get
\begin{equation} 
d^* = d + \frac{a}{4} - \frac{1}{(2 \pi)^{3/2} n^{1/2}}
    + \frac{d}{64 \sqrt{2 \pi^5} (n d^2)^{3/2}}\,.
\label{eqn:dstarhigh}
\end{equation} 
The QC is represented by the last three terms in this equation. They account for, respectively, the kinetic energy of the Fermi sea, the exchange energy, and the correction to the exchange due to screening by the gate. Because of the small numerical factor, the last of these becomes important only at $n^{1/2} d \lesssim 0.25$ [see Fig.~\ref{fig:dstar}(a)].

Finally, the regime of strong correlations induced by the $1/r^3$ tail of the interaction~\cite{Widom1988cea, Skinner2010alc} is realized at intermediate densities, $1 / b^2 \ll n \ll 1 / d ^2$. The energy of such a state is largely insensitive to the spin polarization. Indeed, our calculation shows that the relative difference of $d^*_0$ and $d^*_1$ is small at such $n$. On the other hand, this difference becomes significant in the low-density limit, where the correlations weaken, cf.~Fig.~\ref{fig:dstar}(b).

\textit{Variational PDF.\/} \;
Our calculation also provides the PDF $g_{\text{var}}(r)$ of the screened electron gas [see Fig.~\ref{fig:rscompare}(b)], which has a number of merits. Our PDF is strictly positive, unlike those of some approximate many-body theories~\cite{Giuliani2005qto}. One can show that this $g_{\text{var}}(r)$ and the variational energy components satisfy the virial theorem:
\begin{equation}
\frac{D}{n}\, P = 2 E^{(\text{kin})} - \frac{n}{2} \int d^D r g(r) r \frac{d V}{d r}\, ,
\label{eqn:virial}
\end{equation}
where $P = [\mu(n) - E(n)] n$ is the pressure and $D$ is the space dimension. (The subscript ``var'' in $P$ and $g$ is implicit.) The variational estimates of $E$ and $\mu$, which determine the left-hand side of Eq.~\eqref{eqn:virial}, should be reliable. Therefore, at intermediate values of $r$, which dominate the value of the integral on the right-hand side, our PDF may be a good approximation. At $r$ much larger or smaller than the mean electron spacing, it is probably less accurate. Thus, we have $g_{\text{var}}^\prime(0) / g_{\text{var}}(0) = \frac{2}{D - 1}\, ({\rho_s}/{r_s})$ instead of the exact cusp condition~\cite{Giuliani2005qto} $g^\prime(0) / g(0) = \frac{2}{D - 1}$.

\begin{figure}[t]
\centering
\includegraphics[width=0.35 \textwidth]{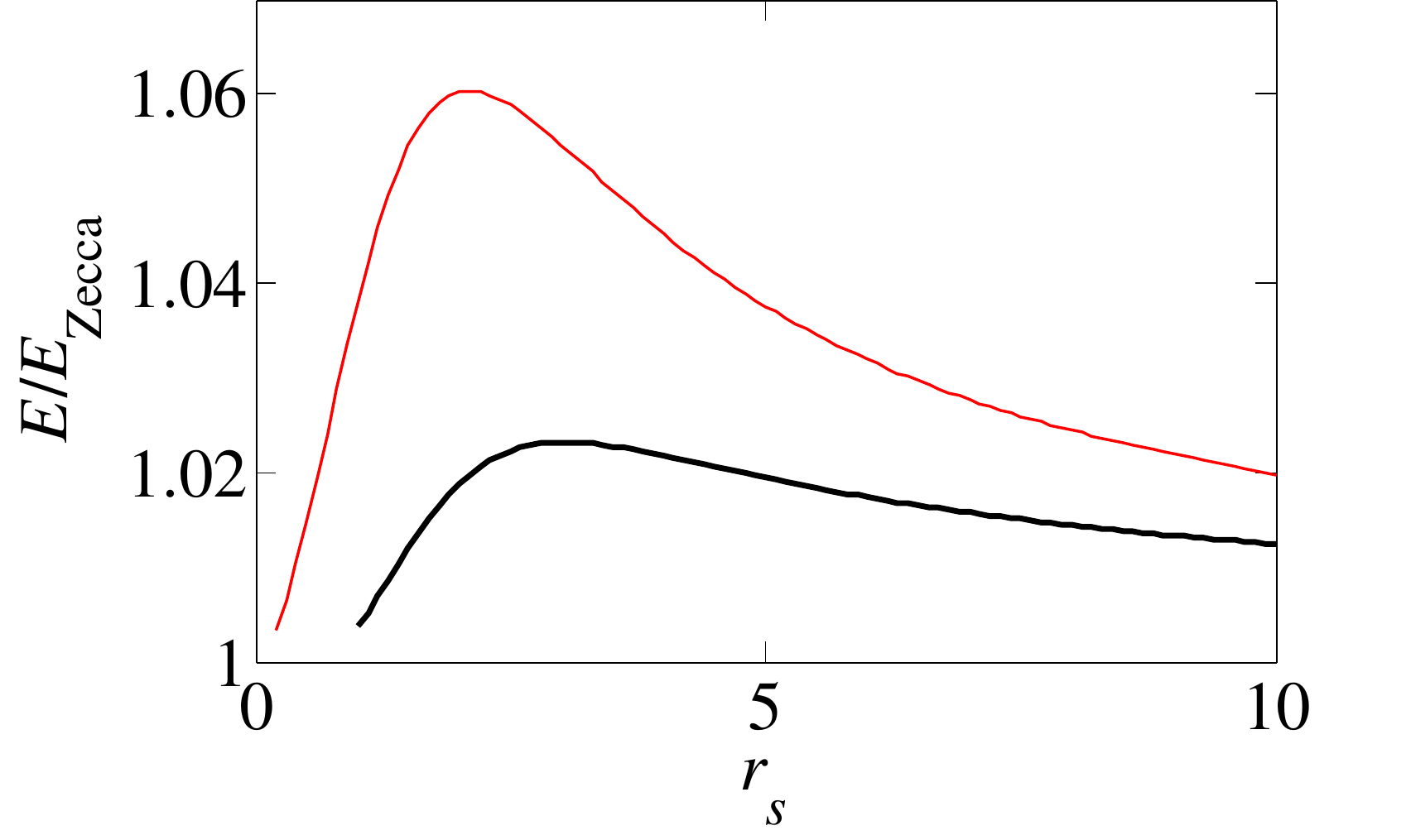}
\caption{(Color online) The ratio of the total variational energy per particle $E$ and the MC energy $E_{\text{Zecca}}$~\cite{Zecca2004ldf} as a function of $r_s$ for a 3D electron gas with the interaction law $V(r) = (e^2 / r)\, \text{erfc}(\mu_c r)$ and $\mu_c = 0.5 / a$ (thick black line). The thin red line shows the HF result.} 
\label{fig:3DEG}
\end{figure}

\textit{Test on a 3D model. \/}
The accuracy of our method is best verified by comparison with MC simulations; however, they are not available for our 2D problem. Instead we did such a test for a 3D electron gas with the interaction potential $V(r) = (e^2 / r)\, \text{erfc}(\mu_c r)$~\cite{Zecca2004ldf} (relevant for mixed DFT schemes). The necessary correlation energies and PDF for the standard 3D gas were taken from Ref.~\onlinecite{Gori-Giorgi2000ass}. The results for the cutoff parameter $\mu_c = 0.5 / a$ are shown in Fig.~\ref{fig:3DEG}. The largest difference between our variational estimate and the MC results~\cite{Zecca2004ldf} for the total energy per particle is $\sim 2\%$, which is a significant improvement over the HF approximation. This example also illustrates the capability of our method to treat other dimensions and/or interaction laws~\cite*{[{A similar variational method is also successful in 1D, see }]Fogler2005gse}.

\textit{Discussion.\/} \;
Our results for $d^*$ can be verified experimentally by measurements of the differential capacitance between an electron gas and a metal gate.  Previously, structures with a distant gate, $n d^2 > 1$, have been studied. At low densities, $d^*$ was seen to be slightly smaller than $d$~\cite{Kravchenko1990eft, Eisenstein1994cot, Dultz2000tso, Allison2006tdo}. In fact, this negative ``QC'' correction arises largely from correlations of classical nature~\cite{Bello1981dol, Skinner2010alc}.

Our theory enables us to compute the capacitance of gated nanostructures in which quantum and correlation effects are not mere corrections. Our results agree well with the data of Ref.~\onlinecite{Li2010lce}. Rigorous testing would require additional experiments at still lower densities and/or a strong in-plane magnetic field (to check the predicted spin dependence). Additionally, the energy per particle we compute here is relevant for understanding impurity screening and therefore transport properties of correlated 2D electron liquids near metallic gates~\cite{Spivak2006tit, Huang2007dom, Ho2008eos, Tracy2009oop, Ho2009gps}.
%
%
Finally, an intriguing direction for future work is to devise a variational method suitable for ultrathin gated structures of graphene~\cite{Xia2009mot, Young2010eco, Ponomarenko2010dos, Droescher2010qca, Henriksen2010mot}, which is an electron system with a nonparabolic spectrum.
 
This paper is supported by the NSF (B.~S.) and by NSF Grant NSF DMR-0706654 (M.~M.~F.). We are grateful to B.~I.~Shklovskii for valuable comments and to the KITP at UCSB for hospitality. 

\bibliography{2DEG-C-variational}
\end{document}